\documentstyle[preprint,aps]{revtex}
\begin{document}
\draft
\title{Static and dynamic properties of frictional phenomena \\
in a one-dimensional system with randomness}
\author{Takaaki Kawaguchi}
\address{Department of Technology, Faculty of Education, Shimane University}
\author{Hiroshi Matsukawa}
\address{Department of Physics, Graduate School of Science, Osaka University}
\date{\today}
\maketitle
\begin{abstract}

Static and dynamic frictional phenomena at the interface with random impurities
are investigated in a two-chain model with incommensurate structure.
Static frictional force is caused by the impurity pinning
and/or by the pinning due to the regular potential, which is responsible for 
the breaking of analyticity transition for impurity-free cases.
It is confirmed that the static frictional force is always finite in the
presence of impurities, in contrast to the impurity-free system.
The nature of impurity pinning is discussed in connection with that
in density waves.
The kinetic frictional force of a steady sliding state is also
investigated numerically.
The relationship between the sliding velocity dependence
of the kinetic frictional force and the strength of impurity
potential is discussed.

\end{abstract}


\vfill\eject
\narrowtext
\section{Introduction}

Friction is an old but unsettled problem in physics\cite{bow}. 
The Coulomb-Amonton's law of friction is one of the most well-known laws in physics.
It has been examined that the law holds well under usual conditions, but breaks 
in some cases.
It is still unsuccessfull, however, to clarify the conditions of Coulomb-Amonton's law 
to hold and the physical mechanism of friction.
On the other hand, recent experimental measurement techniques of frictional phenomena 
have made great progress. One of the direction of such studies of friction tends to 
investigate nano-scale frictional phenomena. For example, atomic and frictional force 
microscope measurements at nanometer scale and quartz-microbalance measurements of 
frictional phenomena of monolayers have been performed  extensively\cite{binnig,krim1,per}.
Stimulated by these experiments, theoretical approaches based on a microscopic model have 
been investigated extensively \cite{sliding}.

As a simple microscopic model of friction, the Frenkel-Kontorova model (FK model)\cite{fk} 
has been employed in several works\cite{aubry1,hira1,shinjo,etc}. 
The FK model consists of atoms interacting with each other via a harmonic force under 
a periodic potential, which is originally sinusoidal. 
In the case that the ratio between the mean atomic spacing and the period of 
the potential is irrational, i.e., in an incommensurate case, the FK model shows 
a phase transition that is called the breaking of analyticity transition or 
Aubry transition\cite{aubry1}. 
When the amplitude of sinusoidal potential is smaller than a certain critical value, 
the maximum static frictional force vanishes. Otherwise it becomes finite.
The existence of this phase transition was originally pointed out in a one-dimensional 
incommensurate FK model by Aubry\cite{aubry1}. 
In three-dimensions, the peculiar frictional nature of the incommensurate FK model 
was found by Hirano and Shinjo\cite{hira1,shinjo}. They pointed out the possibility of 
the appearance of a superlubrication state, that is the vanishing static frictional 
force state, between incommensurate surfaces in real materials.  
Matsukawa and Fukuyama investigated the static and kinetic properties of friction 
in a one-dimensional model\cite{matsu1}. The model employed in their study consists 
of two atomic chains, where interchain force works between atoms in one chain and 
those in another and harmonic force works between neighber atoms in each chain.
The effect of energy dissipation and the relaxation of atoms in both chains are 
taken into accout properly.
On the basis of their model, a definitive expression of both static and 
kinetic frictional force can be derived. They found that the critical value of 
the potential strength of the breaking of analyticity transition 
depends strongly on the rigidity of the chains. In the case that atoms in the lower chain
are fixed periodically, which corresponds to the FK model, that critical value is large. 
When atoms in the lower chain can relax, that critical value decreases and 
the parameter range for the state with vanishing maximum static frictional force 
becomes small.
They also found that the sliding velocity dependence of the kinetic frictional force 
becomes weaker as the maximum static frictional force increases.

In these theoretical studies of friction, surface randomness has not been taken 
into account.
In a realistic system, however, there always exists surface randomness. 
The effects of surface randomness should be taken into consideration to understand 
frictional phenomena in realistic systems. 
Based on a variational method and a perturbational analysis, Matsukawa and Fukuyama 
showed that the state with vanishing maximum static frictional force becomes unstable 
against surface randomness\cite{matsu2}.
The frictional transition is smeared out because the randomness breaks 
the translational symmetry of the system. Hence a finite maximum static frictional force 
appears even if the interatomic interaction is sufficiently small \cite{robin}.

While finite static frictional force is expected to appear in systems 
with surface randomness, 
the magnitude of the maximum static frictional force has not been clarified yet. 
Furthermore, surface randomness influences the kinetic friction. The velocity dependence 
of the kinetic frictional force would be different from that of the system 
with clean surfaces.
In this paper, using a one-dimensional incommensurate model, we investigate the static 
and kinetic frictional forces subject to surface randomness.

This paper is organized as follows. In section {\bf II} we define a one-dimensional 
incommensurate model of friction and introduce surface randomness. 
Based on a numerical study and the impurity pinning theory of density waves, 
we discuss the static frictional force in section {\bf III}. 
The kinetic frictional force, in particular its velocity dependence, 
is investigated in section {\bf IV}. Section {\bf V} is devoted to summary and discussion.
\section{Theoretical model of friction}
In order to calculate frictional force, we consider a one-dimensional model of friction 
proposed by Matsukawa and Fukuyama\cite{matsu1}. 
This model can deal with the friction between two deformable atomic chains interacting 
each other. 
To simplify the problem, however, we here assume that one of the two chains is fixed and 
the other chain can be deformed.  
The atoms of the fixed chain are periodically arranged at a regular position.
We call the deformable chain the upper body and the fixed chain the lower body. 
The atoms in the upper body have a one-dimensional degree of freedom in the direction 
parallel to the chain. 
The interatomic force of the upper body is approximated by the harmonic one.
The effect of the energy dissipation is taken into account by the term in the equation 
of motion which is proportional to the difference between the velocity of each atom and 
that of the center of gravity. 
The external force is applied to the upper body to the horizontal direction. 
Assuming an overdamped condition, we get the equation of motion of the i-th atom of 
the upper body as follows.
\begin{eqnarray}
\lefteqn{
m_a\gamma_a\{ \dot{u}_i-\langle\dot{u}_i\rangle_i\} } \nonumber \\
&=&  K_a\{ u_{i+1}+u_{i-1}-2u_i\} +\sum_{j\in b} F_I(u_i-v_j) +F_{ex} ,  \end{eqnarray}
where $u_i$ ($v_i$) is the position of the i-th atom of the upper (lower) body, 
$m$ the atomic mass, $\gamma$ the parameter of energy dissipation, 
$K_a$ the strength of the interatomic froce, and $\langle\dot{  u}_i\rangle_i$ 
represents the average of $\dot{u}_i$ with respect to $i$.
$F_I$ and $F_{ex}$ are the interchain force between atoms in different chains 
and the external force, respectively.
We adopt the following interchain potential:
\begin{equation}
{U_{I}}=-\frac{K_I}{2}\exp{\left[-4\left(\frac{x}{c_b}\right)^2\right]}, 
\end{equation}
where $K_I$ is the interchain potential strength and 
$c_b$ the mean atomic spacing of the lower body. 
The interchain force is given by $F_I(x)=-\frac{d}{dx}{U_{I}}$.
Since we now consider the case where all the atoms of the lower body are fixed, 
the total interchain force that acts on the i-th atom of the upper body is 
a function of $u_i$ only, which is obtained by summing up the contribution 
from all the atoms of the lower body.  
\begin{equation}
\sum_{j\in a} F_I(u_i-v_j) = \tilde{F}_I(u_i).
\end{equation}
This means that the present model corresponds to the kinetic FK model 
with energy dissipation. 

Now we consider surface randomness such as impurities. 
It is assumed that the interchain potential consists of two parts, i.e., 
the original regular potential and the impurity potential, 
which is distributed randomly.
Furthermore we assume that the impurity potential has the same form as the original one 
except for the potential strength. Hence the total interchain potential is given by
\begin{equation}
{U_{I}}^T (u_i)=\sum_{j=1}^{N_b} {U_{I}} (u_i-v_j) 
+ \sum_{j= {\rm impurity\; sites}}^{N_I} {U_{I}}^{\rm imp} (u_i-v_j),
\end{equation}
where $U_I$ represents the original interatomic potential, 
${U_{I}}^{\rm imp}$ the impurity potential and $N_I$ is the number of impurities. 
The impurity potential, ${U_{I}}^{\rm imp}$, is written as
\begin{equation}
{U_{I}^{\rm imp}}(x)= -\frac{K_{I}^{\rm imp}}{2} 
\exp{\left[-4\left(\frac{x}{c_b}\right)^2\right]},
\end{equation}
where $K_I^{\rm imp}$ is the strength of the impurity potential. 
In this model, the frictional force $F_{\rm fric}$ in a steady state can be calculated 
by the following equation: 
 {$\displaystyle  F_{\rm fric} =- \sum_{i\in a}\left\langle F_I( u_i) 
\right\rangle_{t}, $}
where $\left\langle \dots \right\rangle_{t} $ means the temporal average and 
$F_I(x)=-\frac{d}{dx}{U_{I}}^T (x)$.
Because the system is in a steady state $F_{\rm fric}$ is equal to 
the total external force, $N_a F_{ex}$.  

The equation of motion is numerically solved using the Runge-Kutta formula. 
In the calculation the periodic boundary condition, $u_i=u_{i+N}+L$, is adopted,  
where $L$ is the system size. 
Then the ratio $c_a/c_b$ is equal to the ratio $N_b/N_a$, where $c_a$ is 
the mean lattice spacing of the upper body and $N_a$ and $N_b$ are the number of atoms 
of the upper and lower bodies, respectively.

We here consider an incommensurate system with periodic boundary condition and 
determine the ratio $c_a/c_b$ based on the continued-fraction expansion of the golden mean.
\begin{equation}
\frac{c_a}{c_b}=\frac{N_b}{N_a}=\frac{144}{89}=1.6179\cdots,
\end{equation}
where the expansion is truncated at the tenth order.
The parameters of the present model are set as
\begin{eqnarray}
c_a=\frac{144}{89}, c_b=1, m= K_a =\gamma=1, N_I=70.
\end{eqnarray}  
The number of impurities $N_I$ is fixed throughout the present numerical simulation. 
With increasing external force, the maximum static frictional force, i.e., 
the threshold force at which the upper body starts to slide, is numerically estimated. 
The kinetic frictional force is evaluated in steady sliding states 
with constant velocities \cite{matsu1}.
\section{Maximum static frictional force and impurity pinning effects}

For convenience of the following discussion, we define some energy scales 
that characterize the present model. 
The impurity potential energy, the elastic energy, and the coupling energy to 
the external force per impurity are expressed as $E_i$, $E_{el}$, and $E_f$, respectively, 
and are given by 
\begin{eqnarray}
E_i &=& \frac{1}{2}{K_{I}}^{\rm imp},\\
E_{el} &=& \frac{1}{2}{K_{a}} \times N_I / N_b,\\
E_f &=& F_{ex}\times N_b / N_I.
\end{eqnarray}
For the comparison between these energy scales it is convenient to introduce 
the parameters $\varepsilon_i$ and $\varepsilon_f$ as follows: 
\begin{eqnarray}
\varepsilon_i &=& E_i / E_{el}, \\
\varepsilon_f &=& E_f / E_{el}.
\end{eqnarray}
The strengths of the impurity potential and the external force are discussed 
in terms of $\varepsilon_i$ and $\varepsilon_f$, respectively. 
According to the impurity pinning theory of density waves\cite{matsu3,fuku1,fuku2} 
the nature of the impurity pinning is quite different whether $\varepsilon_i \gg 1$ 
or $\varepsilon_i \ll 1$.

We now attempt to interpret the impurity pinning effect as the origin of static friction. 
The scenario is as follows.
For large $\varepsilon_i$ the effect of the impurity potential overcomes those of 
the elastic energy, and strong pinning states, in which each impurity pins the chain 
and the large lattice distortion is accompanied, are preferred. 
Maximum static frictional force is expected to be proportional to the pinning energy, 
which is the sum of the impurity potential energy and the elastic energy 
in the pinned state.
Hence, the normalized maximum static frictional force $\varepsilon_{f}^{\rm m.s.}$ 
behaves as 
\begin{equation}
\varepsilon_{f}^{\rm m.s.} \propto \varepsilon_{i}.
\end{equation}
On the other hand, for small values of $\varepsilon_i$
each impurity cannot pin the upper body due to the resulting large loss of elastic energy. 
Instead collection of impurities in a finite domain pin the upper body. 
The size of the domain is determined so as to minimize the pinning energy. 
In each domain the system can gain the impurity energy in the scale of fluctuation.
Then we get
\begin{equation}
\varepsilon_{f}^{\rm m.s.} \propto {\varepsilon_{i}}^{\frac{4}{3}},
\end{equation}
in the case of one-dimensions.

The consideration above, however, does not take into accout the effect of the originally 
underlying regular potential $U_{I}$. 
Even for impurity-free cases there can exist the pinning behavior in the present system, 
i.e., the appearance of the finite maximum static frictional force due to 
the Aubry transition for $K_I$ larger than a certain critical value. 
This effect is crucial in considering the static frictional force.

Figs. 1(a)-(c) show $\varepsilon_f^{\rm m.s.}$ versus $\varepsilon_i$ curves 
for the several strengths of the interchain potential $K_I$, 
where the maximum static frictional forces obtained are averaged over 3 to 10 samples.
For $K_I=0.05$ (Fig. 1(a)), the weak impurity pinning relation ($\varepsilon_{f}^{\rm m.s.} 
\propto {\varepsilon_{i}}^{\frac{4}{3}}$) and the strong impurity pinning relation 
($\varepsilon_{f}^{\rm m.s.} \propto \varepsilon_{i}$) hold well 
in the regimes $\varepsilon_i \leq 10$ and $\varepsilon_i > 20$, respectively. 
There exists a  crossover point between the two characteristic impurity pinning states 
around $\varepsilon_i \sim 20$.
Additional crossover behavior appears for $K_I=1$ where the corresponing clean systems have 
finite static frictional force. As observed in Fig. 1(b), the crossover between weak and 
strong impurity pinning states found for $K_I=0.05$ apparently exist around 
$\varepsilon_{i} \sim 20$ even for $K_I=1$. 
The regime of the weak impurity pinning states for $K_I=1$, however,  
is rather narrowed, and apparent deviation from the weak impurity pinning behavior 
is observed for $\varepsilon_i \leq 2$. 
This deviational behavior is considered a crossover between the weak impurity pinning state 
and an another state.
The latter is the state pinned mainly by the regular potential and 
is close to the breaking of analyticity state due to 
the Aubry phase transition in the impurity-free case.
In this regime  
the impurity potential is considered approximately to be a weak perturbation to 
the impurity-free state.
However, in the regimes $\varepsilon_i > 2$, the effect of the impurity potential 
becomes dominant, and then the weak pinning state appears.
The deviation point from the weak pinning state shifts to the large  $\varepsilon_i$ regime 
as $K_I$ increases. 
For a strong interaction ($K_I=6$, Fig. 1(c) ), the weak pinning regime is not observed. 
It is confirmed from the fact that the exponent of $\varepsilon_f^{\rm m.s.}$ 
on $\varepsilon_i$ 
in the regime $\varepsilon_i \leq 10$ is quite different from that of the weak pinning state.
The weak pinning states are destroyed by the regular interchain potential 
because this potential always dominates the impurity potential 
in the weak impurity pinning regimes 
$\varepsilon_i \leq 10$. 

\section{Kinetic frictional force}

In this section we investigate the effect of the impurity potential on 
the velocity dependence of the kinetic frictional force.
Figs.2(a-b) show the velocity($v$) dependence of the kinetic frictional force 
for various strengths of the impurity potential. These values of the impurity potential 
correspond to the weak pinning regime. 
In addition to the finite $\varepsilon_i$ cases we show the kinetic frictional force of 
the clean system($\varepsilon_i=0$) for comparison. 
In a weak regular interchain interaction case, $K_I=0.05$(Fig.2(a)), 
it is obviously observed that the kinetic frictional force for $\varepsilon_i=0$ 
exhibits velocity strengthening($F_{\rm fric}\propto v$) and 
weakening($F_{\rm fric}\propto v^{-1}$) features in low and high velocity regimes, 
respectively. 
Such behavior is excellently explained by a perturbational analysis \cite{soko,kawa1}. 
In the case of finite $\varepsilon_i$, the behavior is quite different and 
the frictional force tends to a finite value in the limit of vanishing velocity \cite{caroli}. 
It is an effect of the impurity potential. 
For small $\varepsilon_i$ the effect is more significant at low velocities 
than at high velocities, 
and additional velocity-weakening kinetic frictional force appears 
in the low volocity regime.
As the strength of the impurity potential increases the maximum static frictional force 
becomes larger than the kinetic frictional force. 
We here note that the behavior of kinetic frictional force 
in disordered systems 
was discussed theoretically by Sokoloff \cite{soko2}. 
When applying his method to one-dimensions, a velocity-weakening feature of 
kinetic frictional force can be deduced 
based on the perturbation theory for a general disordered potential.
The result is, however, obtained in the underdamped regime, while the present
results are obtained in the overdamped regime.
Fig.2(b) shows the result on the kinetic frictional force for $K_I=1$ 
where there exists large maximum static frictional force. 
In this case the kinetic frictional force for $\varepsilon_i=0$ shows a similar velocity 
dependence to that for finite $\varepsilon_i$, 
that is, almost velocity-independent behavior in the low-velocity regime and inversely 
proportional behavior to the sliding velocity in the high-velocity regime.
In both cases of Figs.2(a) and (b) the strength of the kinetic frictional force 
monotonically increases with increasing $\varepsilon_i$.

The kinetic frictional force in the sliding states evolved 
from strong impurity pinning states is shown in Fig.3. 
Under this condition, a very large maximum static frictional force appears 
and the static and kinetic friction is mainly caused by the interaction 
between the chain and the impurities. The kinetic frictional force in Fig.3 is similar to 
that in Fig.2(b), i.e., the velocity-independent and velocity-weakening features.
No distinct qualitative difference in the behavior of the kinetic frictional force is 
observed between the different strengths of the interchain potential, $K_I=0.05$ and $1$. 

From the above results on kinetic frictional force, we found that 
the velocity-weakening behavior in the high-velocity regime is a common feature 
of the present model 
and the velocity-independent behavior in the low-velocity regime is peculiar to 
the system with large maximum static frictional force.

\section{Summary and discussion}

We have calculated the static frictional force by computer simulations 
using a one-dimensional incommensurate model of friction. 
It has been numerically confirmed that the randomness due to impurities destroys 
the vanishing static frictional force states and the maximum static frictional force always 
becomes finite. 
This is consistent with the variational calculation by Matsukawa and Fukuyama \cite{matsu2}.

We have also found that frictional phenomena have a close relationship to 
impurity pinning phenomena in density wave systems.
For the case of weak regular interchain interaction the nature of impurity pinning states 
changes from weak pinning to strong pinning as the strength of impurity potential increases.
Similar behavior is also observed in density waves\cite{matsu3,brill}.
For a strong regular interchain interaction the maximum static frictional force is finite 
even in impurity-free systems. The feature of the impurity pinning in the weak pinning regimes 
is modified. In this case the pinning is caused essentially by the strong regular potential 
rather than by the impurity potential. 
The crossover behavior of the maximum static frictional force has been found from 
the pinned sate affected by the regular potential to the weak or strong impurity pinning state 
as the impurity potential is strengthened.

We have also investigated the kinetic frictional force in the presence of impurities. 
It has been found that surface randomness is crucial to the velocity dependence of 
the kinetic frictional force. 
The behavior of the kinetic frictional force in the present work should be compared with 
that of the Coulomb-Amonton's law, i.e., the kinetic frictional force is less than 
the maximum static frictional force and does not depend on the sliding velocity. 
In the case of large maximum static frictional force the velocity-independent behavior 
of kinetic frictional force is observed in the low-velocity regime.
The magnitude of the kinetic frictional force is, however, equal to 
the maximum static frictional force.

In a realistic system there are several complicated factors that are not taken into 
consideration in the present theoretical model. 
For example, the effect of other types of randomness such as steps on surfaces is not 
examined here.
Such an effect may be also an important factor, but we expect 
that the velocity dependence of the kinetic frictional force 
would not depend on the details of the randomness and 
would show a similar behavior to that of the present model 
because the major contribution to the kinetic frictional force comes from 
the excitation of phonons with long wavelength.
Nevertheless several important issues are left unsolved and therefore 
we need further investigation to compare 
the detailed features of the friction model with those of realistic systems.

\acknowledgments
We would like to thank Professor H. Fukuyama, Professor M. Robbins, and Dr. M. Hirano 
for valuable discussions.

\appendix
\section*{}
 We make some comments on Eqs. (13) and (14). 
As mentioned in the text, these relations are derived using the impurity pinning theory 
of density waves. Concerning the analogy between the present model and the model of 
density waves, the following points should be noted.
In the case of density waves \cite{fuku2}, the average impurity energy vanishes 
when density waves are rigid and have perfect periodicity. 
The energy gain due to the impurity potential arises from the distortion of density waves.
On the other hand, since the present model has the attractive potential $U_I^{\rm imp}$, 
the energy gain always exists even without the distortion of the chain. 
This energy, however, does not affect the frictional force because the amount of the energy 
is independent of the position of the chain.
In both systems the energy cost caused by the distortion is relevant to the pinning effect. 
Hence, in the present model, the same behavior of the impurity pinning as that 
in density waves is observed.

Fig. 1. {$\varepsilon_i$ versus $\varepsilon_f^{\rm m.s.}$. 
The solid and dashed lines represent the relations: $\varepsilon_f^{\rm m.s.} 
\propto {\varepsilon_{i}}^{\frac{4}{3}}$ and $\varepsilon_f^{\rm m.s.} \propto 
\varepsilon_{i}$}, respectively. 
When $\varepsilon_{i}=0$($K_I^{\rm imp}=0$), $\varepsilon_f^{\rm m.s.}= 0, 0.760$, 
and $18.0$ for $K_I= 0.05, 1$, and $6$ respectively.\\

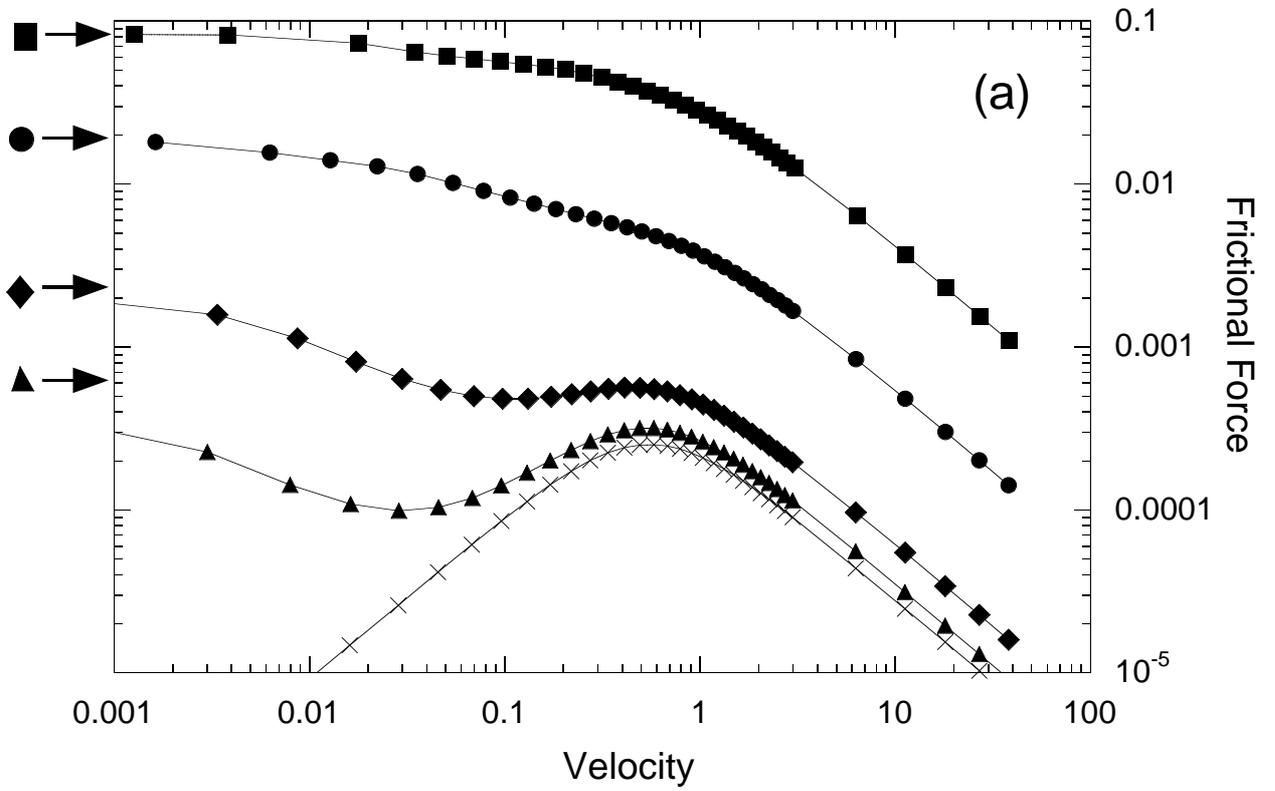
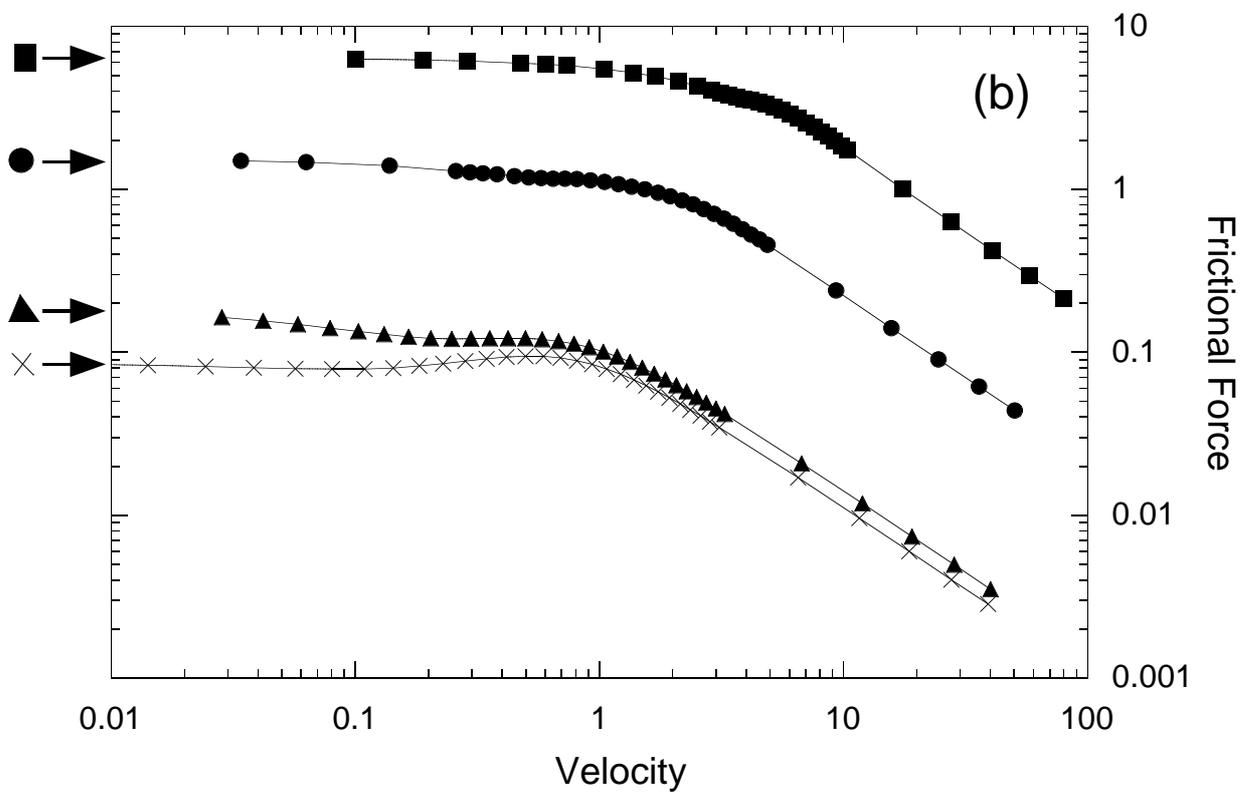
Fig. 2. {The velocity dependence of kinetic frictional force for weak pinning. 
(a) $K_I=0.05$. Squares, circles, diamonds, triangles, and crosses represent 
the results for $\varepsilon_i= 0.926, 0.309, 0.0617, 0.0206,$ and 
$0$ ($K_I^{\rm imp}=0.45, 0.15, 0.03, 0.01,$ and $0$), respectively. (b) $K_I=1$.  
Squares, circles, triangles and crosses represent the results for 
$\varepsilon_i= 18.5, 6.17, 0.411,$ and $0$ ($K_I^{\rm imp}=9, 3, 0.2,$ and $0$), 
respectively.
Arrows indicate the maximum static frictional force.
}\\

Fig. 3. {The velocity dependence of kinetic frictional force for strong pinning. 
Circles and triangles represent the results for 
$\varepsilon_i=51.3$ ($K_I^{\rm imp}=24.95$) and $K_I=0.05$ and 
for $\varepsilon_i=90.5$ ($K_I^{\rm imp}=44$) and $K_I=1$, respectively.
Arrows indicate the maximum static frictional force.
 }

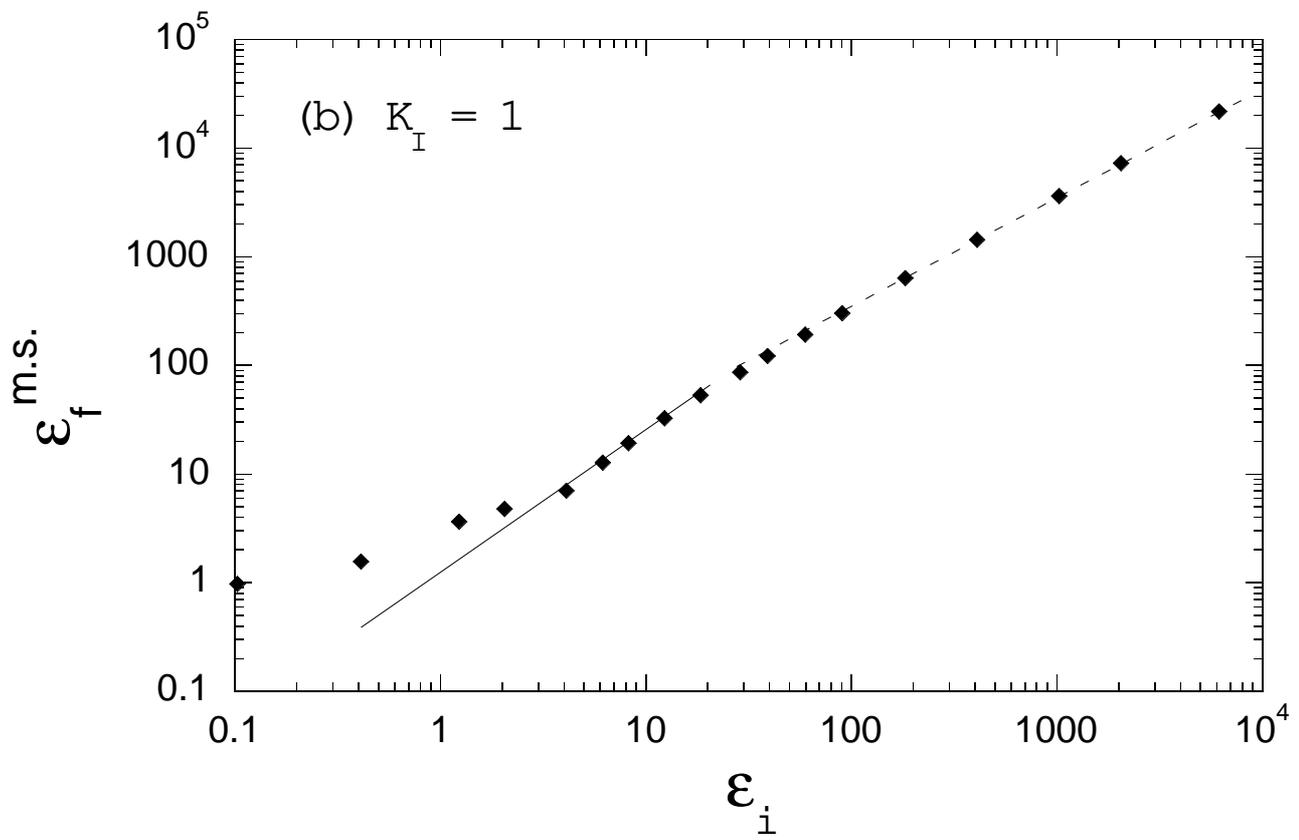
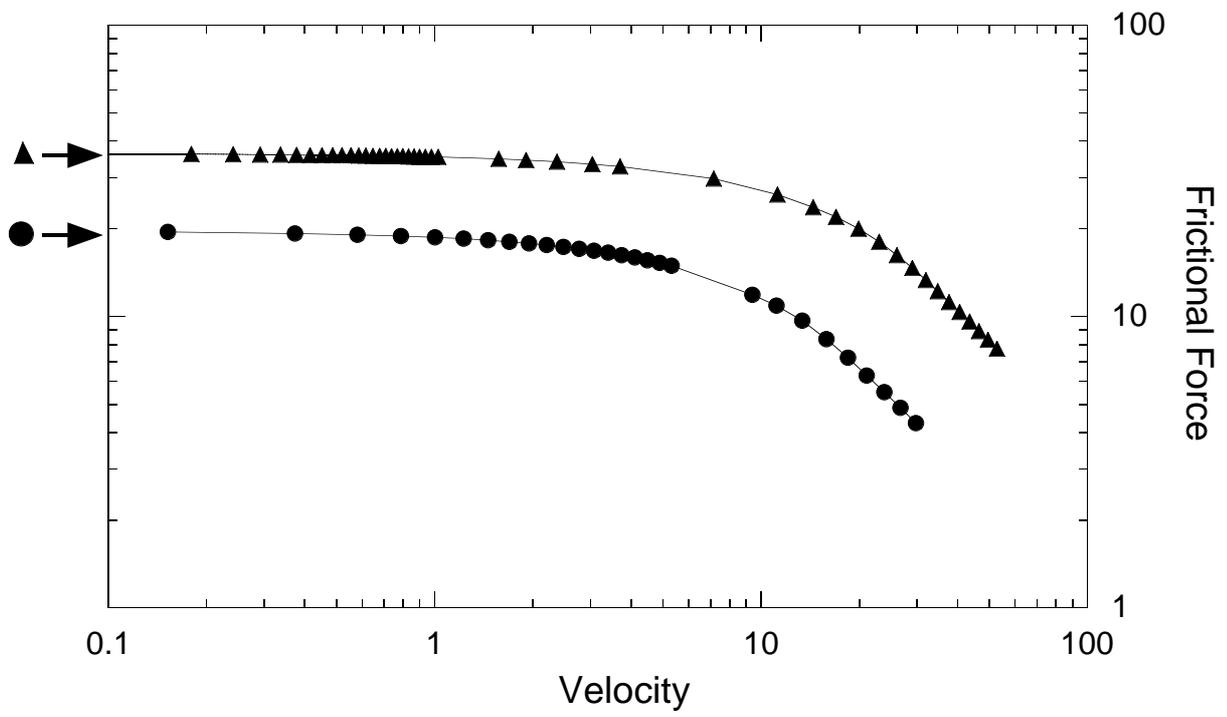
\end{document}